\documentclass[journal]{IEEEtran}
\usepackage{cite}

\usepackage{graphicx}
\DeclareGraphicsExtensions{.pdf,.jpeg,.png,.jpg, .svg}

\usepackage{amssymb}
\usepackage[cmex10]{amsmath}
\usepackage{array}

\usepackage[font = scriptsize]{caption}
\usepackage{color}
\usepackage{commath}
\usepackage{amssymb}
\usepackage{lipsum,graphicx,multirow}
\usepackage{bm}
\usepackage{amsmath}
\usepackage{subcaption}
\usepackage[labelformat=simple, font=scriptsize]{subcaption}

\usepackage{mathtools}

\newcommand{\ie}{\textit{i}.\textit{e}. }

\usepackage[utf8]{inputenc}
\usepackage{algorithm}

\usepackage[]{algpseudocode}

\algnewcommand\algorithmicinput{\textbf{Input:}}
\algnewcommand\Input{\item[\algorithmicinput]}
\algnewcommand\algorithmicoutput{\textbf{Output:}}
\algnewcommand\Output{\item[\algorithmicoutput]}
\algnewcommand\algorithmichline{}
\algnewcommand\Hline{\item[\algorithmichline]}
\makeatletter
    \newcommand*{\algrule}[1][\algorithmicindent]{\makebox[#1][l]{\hspace*{.5em}\thealgruleextra\vrule height \thealgruleheight depth \thealgruledepth}}%
\newcommand*{\thealgruleextra}{}
\newcommand*{\thealgruleheight}{.85\baselineskip}
\newcommand*{\thealgruledepth}{.25\baselineskip}

\newcount\ALG@printindent@tempcnta
\def\ALG@printindent{%
    \ifnum \theALG@nested>0
        \ifx\ALG@text\ALG@x@notext
        \else
            \unskip
            \addvspace{-1pt}
            \ALG@printindent@tempcnta=1
            \loop
                \algrule[\csname ALG@ind@\the\ALG@printindent@tempcnta\endcsname]%
                \advance \ALG@printindent@tempcnta 1
            \ifnum \ALG@printindent@tempcnta<\numexpr\theALG@nested+1\relax
            \repeat
        \fi
    \fi
    }%
\usepackage{etoolbox}
\patchcmd{\ALG@doentity}{\noindent\hskip\ALG@tlm}{\ALG@printindent}{}{\errmessage{failed to patch}}
\makeatother

\newbox\statebox
\newcommand{\myState}[1]{%
    \setbox\statebox=\vbox{#1}%
    \edef\thealgruleheight{\dimexpr \the\ht\statebox+1pt\relax}%
    \edef\thealgruledepth{\dimexpr \the\dp\statebox+1pt\relax}%
    \ifdim\thealgruleheight<.75\baselineskip
        \def\thealgruleheight{\dimexpr .75\baselineskip+1pt\relax}%
    \fi
    \ifdim\thealgruledepth<.25\baselineskip
        \def\thealgruledepth{\dimexpr .25\baselineskip+1pt\relax}%
    \fi
    \State #1%
    \def\thealgruleheight{\dimexpr .75\baselineskip+1pt\relax}%
    \def\thealgruledepth{\dimexpr .25\baselineskip+1pt\relax}%
}

\algdef{SE}[DOWHILE]{Do}{doWhile}{\algorithmicdo}[1]{\algorithmicwhile\ #1}%

\begin{document}
\title{Energy-Efficient Transponder Configuration for FMF-based Elastic Optical Networks}
\author{Mohammad~Hadi,~\IEEEmembership{Member,~IEEE,}
        and~Mohammad~Reza~Pakravan,~\IEEEmembership{Member,~IEEE}
}

\maketitle

\begin{abstract}
\boldmath
We propose an energy-efficient procedure for transponder configuration in FMF-based elastic optical networks in which quality of service and physical constraints are guaranteed and joint optimization of transmit optical power, temporal, spatial and spectral variables are addressed. We use geometric convexification techniques to provide convex representations for quality of service, transponder power consumption and transponder configuration problem. Simulation results show that our convex formulation is considerably faster than its mixed-integer nonlinear counterpart and its ability to optimize transmit optical power reduces total transponder power consumption up to $32\%$. We also analyze the effect of mode coupling and number of available modes on power consumption of different network elements.
\end{abstract}
\begin{IEEEkeywords}
Convex optimization, Green communication, Elastic optical networks, Few-mode fibers, Mode coupling.
\end{IEEEkeywords}

\section{Introduction}\label{sec_I}
\IEEEPARstart{3}{D} temporally, spectrally and spatially Elastic Optical Network (EON) has been widely acknowledged as the next generation high capacity transport system and the optical society has focused on its architecture and network resource allocation techniques \cite{proietti20153d}. EONs can provide an energy-efficient network configuration by adaptive 3D resource allocation according to the communication demands and physical conditions. Higher energy efficiency of Orthogonal Frequency Division Multiplex (OFDM) signaling has been reported in \cite{khodakarami2014flexible} which nominates OFDM as the main technology for resource provisioning over 2D resources of time and spectrum. On the other hand, enabling technologies such as Few-Mode Fibers (FMFs) and Multi-Core Fibers (MCFs) have been used to increase network capacity and efficiency through resource allocation over spatial dimension \cite{saridis2015survey, proietti20153d}. Although many variants of algorithms have been proposed for resource allocation in 1D/2D EONs \cite{chatterjee2015routing}, joint assignment of temporal, spectral and spatial resources in 3D EONs needs much research and study. Among the available works on 3D EONs, a few of them have focused on energy-efficiency which is a fundamental requirement of the future optical networks \cite{muhammad2015resource, winzer2011energy, yan2017joint}. Moreover, the available energy-efficient 3D approaches do not consider transmit optical power as an optimization variable which results in inefficient network provisioning \cite{hadi2017energy}.

Flexible resource allocation is an NP-hard problem and it is usually decomposed into several sub-problems with lower complexity \cite{khodakarami2016quality}. Following this approach, we decompose the resource allocation problem into 1) Routing and Ordering Sub-problem (ROS) and 2) Transponder Configuration Sub-problem (TCS) and mainly focus on TCS which is more complex and time-consuming \cite{yan2015resource, hadi2017resource}. We consider FMF because it has simple amplifier structure, easier fusion process, lower nonlinear effects and lower manufacturing cost compared to other Space-Division Multiplexed (SDM) optical fibers \cite{saridis2015survey, ho2013mode}. In TCS, we optimally configure transponder parameters such as  modulation level, number of sub-carriers, coding rate, transmit optical power, number of active modes and central frequency such that total transponders power consumption is minimized while Quality of Service (QoS) and physical constraints are met. Unlike the conventional approach, we provide convex expressions for transponder power consumption and Optical Signal to Noise Ratio (OSNR), as an indicator of QoS. We then use the results to formulate TCS as a convex optimization problem which can efficiently be solved using fast convex optimization algorithms. We consider transmit optical power as an optimization variable and show that it has an important impact on total transponder power consumption. Simulation results show that our convex formulation can be solved almost $20$ times faster than its Mixed-Integer NonLinear Program (MINLP) counterpart. Optimizing transmit optical power also improves total transponder power consumption by a factor of $32\%$ for European Cost239 optical network with aggregate traffic $60$ Tbps. We analyze the effect of mode coupling on power consumption of the different network elements. As simulation results show, total network power consumption can be reduced more than $50\%$ using strongly-coupled FMFs rather than weakly-coupled ones. Numerical outcomes also demonstrate that increasing the number of available modes in FMFs provides a trade-off between FFT and DSP power consumption such that the overall transponder power consumption is a descending function of the number of available modes. 
\section{System Model}\label{Sec_II}
\begin{figure}[t!] 
\center{\includegraphics[scale=0.38]{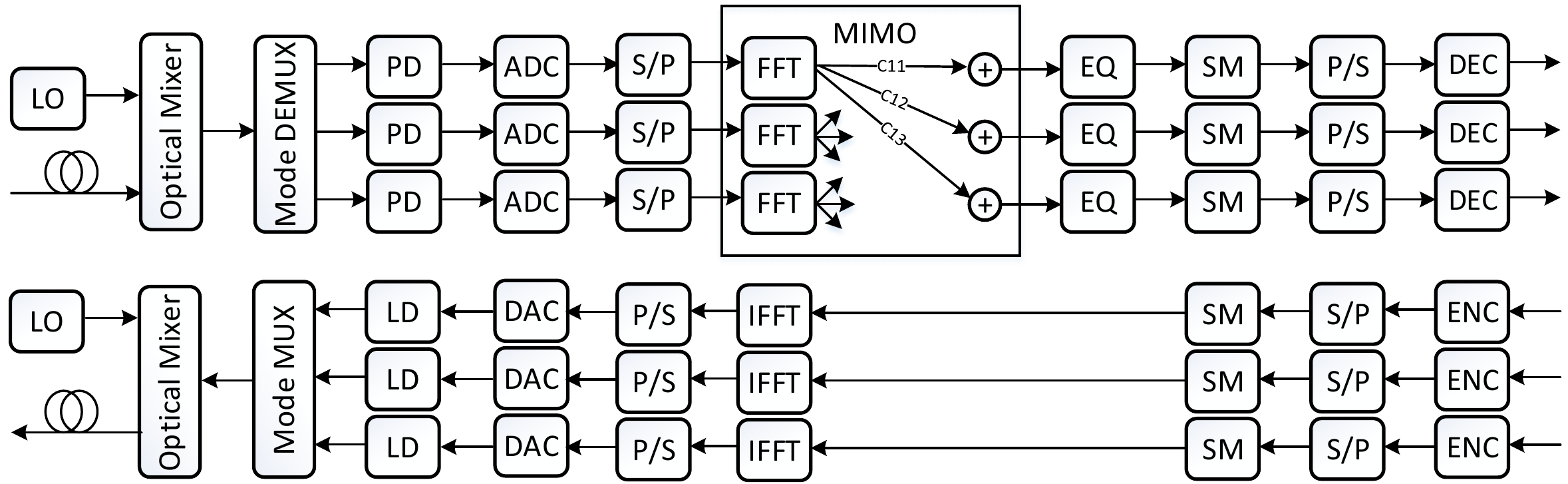}}
\center{\caption{\label{fig:transponder} Block diagram of a pair of OFDM-based transmit and receive transponders with 3 available modes.\vspace{-5mm}}}
\end{figure}

Consider a coherent optical communication network characterized by topology graph $G(\mathbf{V}, \mathbf{L})$ where $\mathbf{V}$ and $\mathbf{L}$ are the sets of optical nodes and directional optical strongly-coupled FMF links, respectively. The optical FMFs have $\mathcal{M}$ modes and gridless bandwidth $\mathcal{B}$. $\mathbf{Q}$ is the set of connection requests and $\mathbf{Q}_l$ shows the set of requests sharing FMF $l$ on their routes. Each request $q$ is assigned a contiguous bandwidth $\Delta_q$ around carrier frequency $\omega_q$ and modulates $m_q$ modes of its available $\mathcal{M}$ modes. The assigned contiguous bandwidth includes $2^{b_q}$ OFDM sub-carriers with sub-carrier space of $\mathcal{F}$ so, $\Delta_q=2^{b_q}\mathcal{F}$. To have a feasible MIMO processing, the remaining unused modes of a request cannot be shared among others \cite{ho2013mode}. We assume that the assigned bandwidths are continuous over their routes to remove the high cost of spectrum/mode conversion \cite{spectrum2017hadi}. Request $q$ passes $\mathcal{N}_q$ fiber spans along its path and has $\mathcal{N}_{q,i}$ shared spans with request $i\neq q$. Each FMF span has fixed length of $\mathcal{L}_{spn}$ and an optical amplifier to compensate for its attenuation.  There are pre-defined modulation levels $c$ and coding rates $r$ where each pair of $(c, r)$ requires minimum OSNR $\Theta(c, r)$ to get a pre-FEC BER value of $1\times 10^{-4}$ \cite{yan2015resource}. Each transponder is given modulation level $c_q$, coding rate $r_q$ and injects optical power $p_q/m_q$ to each active mode of each polarization. Chromatic dispersion and mode coupling signal broadenings are respectively proportional to $\mathcal{N}_q2^{b_q}$ and $m_q^{-0.78}\sqrt{\mathcal{N}_q}$ with coefficients $\sigma=2\pi\abs{\beta_2}\mathcal{F}\mathcal{L}_{spn}$ and $\varrho=5\Delta\beta_1 \sqrt{L_{sec}\mathcal{L}_{spn}}$ where $\beta_2$ is chromatic dispersion factor and $\Delta\beta_1 \sqrt{L_{sec}}$ is the product of rms uncoupled group delay spread and section length \cite{arik2014adaptive}. Transponders add a sufficient cyclic prefix to each OFMD symbol to resolve the signal broadening induced by mode coupling and chromatic dispersion. Transponders have maximum information bit rate $\mathcal{C}$. There is also a guard band $\mathcal{G}$ between any two adjacent requests on a link. Considering the architecture of Fig. \ref{fig:transponder}, the power consumption of each pair of transmit and receive transponders $P_q$ can be calculated as follows:
\small
\begin{align}\label{eq:trx_pow}
P_q = \mathcal{P}_{trb}+2\mathcal{P}_{edc}m_qr_q^{-1}+2m_q2^{b_q}b_q\mathcal{P}_{fft}+2m_q^22^{b_q}\mathcal{P}_{dsp}
\end{align}
\normalsize
where $\mathcal{P}_{trb}$ is transmit and receive transponder bias term, $\mathcal{P}_{edc}$ is the scaling coefficient of encoder and decoder power consumptions, $\mathcal{P}_{fft}$ denotes the power consumption for a two point FFT operation and $\mathcal{P}_{dsp}$ is the power consumption scaling coefficient of the receiver DSP and MIMO operations \cite{ho2013mode, khodakarami2014flexible}. 

To have a green EON, we need a resource allocation algorithm to determine the values of system model variables such that the transponders consume the minimum power while physical constrains are satisfied and desired levels of OSNR are guaranteed. In general, such a problem is modeled as an NP-hard MINLP optimization problem \cite{yan2015resource}. To simplify the problem and provide a fast-achievable near-optimum solution, the resource allocation problem is usually decomposed into two sub-problems: ROS, where the routing and ordering of requests on each link are defined, and 2) TCS, where transponders are configured.  Usually the search for a near optimal solution involves iterations between these two sub-problems. To save this iteration time, it is of great interest to hold the running time of each sub-problem at its minimum value. In this work, we mainly focus on TCS which is the most time-consuming sub-problem and formulate it as a convex problem to benefit from fast convex optimization algorithms. For a complete study of ROS, one can refer to \cite{hadi2017energy, hadi2017resource}.
\section{Transponder Configuartion Problem}\label{Sec_III}
A MINLP formulation for TCS is as follows:
\small
\begin{align}
&\hspace{-1.5mm}\min_{\mathbf{c}, \mathbf{b}, \mathbf{r}, \mathbf{p},\mathbf{m}, \bm{\omega}} \quad \sum\limits_{q \in \mathbf{Q}}P_q\label{eq:nonlinear_g}\\
&\hspace{-1.5mm}\text{s.t.} \quad \Psi_q \geqslant \Theta_q, \forall q \in \mathbf{Q} \label{eq:nonlinear_c1}\\
& \hspace{-1.5mm} \omega_{\Upsilon_{l, j}}+\frac{\Delta_{\Upsilon_{l, j}}+\Delta_{\Upsilon_{l, j+1}}}{2} + \mathcal{G}\leqslant \omega_{\Upsilon_{l, j+1}}, \forall l \in \mathbf{L}, \forall j \in \mathbf{M}_{1}^{\abs{\mathbf{Q}_l}}  \label{eq:nonlinear_c2}\\
&\hspace{-1.5mm}\frac{\Delta_q}{2} \leqslant \omega_q \leqslant \mathcal{B}- \frac{\Delta_q}{2}, \forall q \in \mathbf{Q}\label{eq:nonlinear_c3}\\
& \hspace{-1.5mm}\mathcal{R}_q \leqslant \frac{2\mathcal{F}^{-1}m_qr_q c_q\Delta_q}{\mathcal{F}^{-1}+\sigma\mathcal{N}_{q}2^{b_q}+\varrho m_q^{-0.78}\sqrt{\mathcal{N}_{q}}}, \forall q \in \mathbf{Q} \label{eq:nonlinear_c4}
\end{align}
\normalsize
where $\mathbf{c}$, $\mathbf{b}$, $\mathbf{r}$, $\mathbf{p}$, $\mathbf{m}$ and $\bm{\omega}$ are variable vectors of transponder configuration parameters \ie modulation level, number of sub-carriers, coding rate, transmit optical power, number of active modes and central frequency. $\mathbf{M}_{a}^{b}$ shows the set of integer numbers $\{a, a+1, \cdots, b-1\}$. The goal is to minimize the total transponder power consumption where $P_q$ is obtained using \eqref{eq:trx_pow}. Constraint \eqref{eq:nonlinear_c1} is the QoS constraint that forces OSNR $\Psi_q$ to be greater than its required minimum threshold $\Theta_q$. $\Psi_q$ is a nonlinear function of $\mathbf{b}$, $\mathbf{p}$, $\mathbf{m}$ and $\bm{\omega}$ while the value of $\Theta_q$ is related to $r_q$ and $c_q$ \cite{hadi2017resource, yan2017joint}. Constraint \eqref{eq:nonlinear_c2} is nonoverlapping-guard constraint that prevents two requests from sharing the same frequency spectrum. $\Upsilon_{l, j}$ is a function that shows which request occupies $j$-th assigned spectrum bandwidth on link $l$ and its values are determined by solving ROS \cite{hadi2017resource, yan2017joint}. Constraint \eqref{eq:nonlinear_c3} holds all assigned central frequencies within the acceptable range of the fiber spectrum. The last constraint guarantees that the transponder can convey the input traffic rate $\mathcal{R}_q$ in which wasted cyclic prefix times are considered. 

Generally, this problem is a complex MINLP which is NP-hard and cannot easily be solved in a reasonable time. Therefore, we use geometric convexification techniques to convert this MINLP to a mixed-integer convex optimization problem and then use relaxation method to solve it. To have a convex problem, we first provide a generalized posynomial expression \cite{boyd2007tutorial} for the optimization and then define a variable change to convexify the problem. A posynomial expression for OSNR of a request in 2D EONs has been proposed in \cite{hadi2017resource}. We simply consider each active mode as an independent source of nonlinearity and incoherently add all the interferences \cite{yan2015resource}. Therefore the extended version of the posynomial OSNR expression is:
\small
\begin{align}\label{eq:xci_app}
\hspace{-3 mm}\Psi_q=  \frac{p_q/m_q}{\zeta \mathcal{N}_q\Delta_q+\kappa_1\varsigma \frac{p_q}{m_q}\sum\limits_{\substack{i \in \mathbf{Q}, q \neq i}}m_i(\frac{p_i}{m_i})^2\mathcal{N}_{q,i}/\Delta_i/d_{q,i}}, \forall q \in \mathbf{Q}
\end{align}
\normalsize
where $\kappa_1=0.4343$, $\zeta=(e^{\alpha \mathcal{L}_{spn}}-1)h\nu n_{sp}$ and $\varsigma=\frac{3\gamma^2}{2\alpha\pi\abs{\beta_2}}$. $n_{sp}$ is the spontaneous emission factor, $\nu$ is the light frequency, $h$ is Planck’s constant, $\alpha$ is attenuation coefficient, $\beta_2$ is dispersion factor and $\gamma$ is nonlinear constant. Furthermore,  $d_{q,i}$ is the distance between carrier frequencies $\omega_q$ and $\omega_i$ and equals to $d_{q,i} = \abs{\omega_q-\omega_i}$. We use $\Theta_q \approx r_q^{\kappa_2}(1+\kappa_3 c_q)^{\kappa_4}$ for posynomial curve fitting of OSNR threshold values where $\kappa_2=3.37$, $\kappa_3=0.21$, $\kappa_4=5.73$ \cite{hadi2017energy}. Following the same approach as \cite{hadi2017resource}, we arrive at this new representation of the optimization problem:
\small
\begin{align}
&\hspace{-1.5mm} \min_{\mathbf{c}, \mathbf{b}, \mathbf{r}, \mathbf{p}, \mathbf{m}, \bm{\omega}, \mathbf{t}, \mathbf{d}} \quad \sum\limits_{q \in \mathbf{Q}}P_{q}+\mathcal{K}\sum\limits_{\substack{q,i \in \mathbf{Q}\\q \neq i, \mathcal{N}_{q,i} \neq 0}}d_{q,i}^{-1}\label{eq:gp_1_g}\\
\nonumber &\hspace{-1.5mm} \text{s.t.} \quad  r_{q}^{\kappa_2} t_{q}^{\kappa_4}\Big[\zeta\mathcal{F}\mathcal{N}_{q}m_qp_{q}^{-1}2^{b_{q}}+ \kappa_1\varsigma\mathcal{F}^{-1}\sum\limits_{i \in \mathbf{Q}, i \neq q}\mathcal{N}_{q,i}p_{i}^{2}m_i^{-1}2^{-b_{i}} \\
 & \hspace{-1.5mm}  d_{q,i}^{-1}\Big]\leqslant 1 , \forall q \in \mathbf{Q} \label{eq:gp_1_c1} \\
\nonumber & \hspace{-1.5mm} \omega_{\Upsilon_{l, j}}+0.5\mathcal{F}2^{b_{\Upsilon_{l, j}}} +\mathcal{G}+ 0.5\mathcal{F}2^{b_{\Upsilon_{l, j+1}}}\leqslant  \omega_{\Upsilon_{l, j+1}}, \forall l \in \mathbf{L}\\
& \hspace{-1.5mm} ,\forall j \in \mathbf{M}_{1}^{\abs{\mathbf{Q}_l}} \label{eq:gp_1_c2}\\
& \hspace{-1.5mm} 0.5\mathcal{F}2^{b_q} + \omega_{q} \leqslant \mathcal{B}, \forall q \in \mathbf{Q} \label{eq:gp_1_c3} \\
& \hspace{-1.5mm} 0.5\mathcal{F}2^{b_q} \leqslant \omega_{q}, \forall q \in \mathbf{Q}\label{eq:gp_1_c4}  \\
\nonumber & \hspace{-1.5mm}  0.5\mathcal{R}_q\mathcal{F}^{-1} r_{q}^{-1}  c_{q}^{-1}m_q^{-1} 2^{-b_{q}}+ 0.5 \sigma \mathcal{N}_q \mathcal{R}_{q} m_q^{-1}r_{q}^{-1}  c_{q}^{-1}\\
& \hspace{-1.5mm} +0.5 \varrho \sqrt{\mathcal{N}_q}  \mathcal{R}_qr_{q}^{-1}  c_{q}^{-1}m_q^{-1.78} 2^{-b_{q}}\leqslant 1,  \forall q \in \mathbf{Q}\label{eq:gp_1_c5}  \\
& \hspace{-1.5mm} 1+\kappa_3c_{q} \leqslant t_{q}, \forall q \in \mathbf{Q}  \label{eq:gp_1_c6}  \\
 &  \hspace{-1.5mm} d_{\Upsilon_{l, i},\Upsilon_{l, j}} + \omega_{\Upsilon_{l, j}}\leqslant  \omega_{\Upsilon_{l, i}}, \forall l \in \mathbf{L},\forall j \in \mathbf{M}_{1}^{\abs{\mathbf{Q}_l}},\forall i \in \mathbf{M}_{j+1}^{\abs{\mathbf{Q}_l}+1} \label{eq:gp_1_c7} \\
& \hspace{-1.5mm} d_{\Upsilon_{l, i},\Upsilon_{l, j}}+\omega_{\Upsilon_{l, i}} \leqslant \omega_{\Upsilon_{l, j}} , \forall l \in \mathbf{L},\forall j \in \mathbf{M}_{2}^{\abs{\mathbf{Q}_l}+1}, \forall i \in \mathbf{M}_{1}^{j} \label{eq:gp_1_c8} 
\end{align}
\normalsize
Ignoring constraints \eqref{eq:gp_1_c6}, \eqref{eq:gp_1_c7}, \eqref{eq:gp_1_c8} and the penalty term of the goal function \eqref{eq:gp_1_g}, the above formulation is equivalent geometric program of the previous MINLP in which expressions \eqref{eq:xci_app} and the mentioned posynomial curve fitting have been used for QoS constraint \eqref{eq:gp_1_c1}. Constraints \eqref{eq:gp_1_c6} and \eqref{eq:gp_1_c7} and the penalty term are added to guarantee the implicit equality of $d_{q,i} = \abs{\omega_{q}-\omega_{i}}$ \cite{hadi2017resource}. Constraint \eqref{eq:gp_1_c6} is also needed to convert the generalized posynomial QoS constraint to a valid geometric expression, as explained in \cite{boyd2007tutorial}. Now, consider the following variable change:
\small
\begin{align}\label{eq:vc}
x=e^{X}:x \in \mathbb{R}_{>0} \longrightarrow X \in \mathbb{R}, \forall x \notin \mathbf{b}
\end{align}
\normalsize
Applying this variable change to the goal function (which is the most difficult part of the variable change), we have:
\small
\begin{align}\label{eq:cv_g}
\nonumber & \sum\limits_{q \in \mathbf{Q}} [\mathcal{P}_{trb}+2\mathcal{P}_{edc}e^{m_q-r_q}+5.36e^{0.82b_q+m_q}\mathcal{P}_{fft}+ 2e^{2m_q}2^{b_q}\mathcal{P}_{dsp}]\\
 &  +\mathcal{K}\sum\limits_{\substack{q,i\in \mathbf{Q}, q \neq i, \mathcal{N}_{q,i} \neq 0}}e^{-d_{q,i}}
\end{align}
\normalsize
Clearly, $e^{-d_{q,i}}$, $e^{m_q-r_q}$ and $e^{2m_q}2^{b_q}$ are convex over variable domain.  We use expression $5.36e^{0.82b_q+m_q}$ to provide a convex approximation for the remaining term $2e^{m_q}b_q2^{b_q}$. The approximation relative error is less than $3\%$ for practical values of $m_q \geqslant 1$ and $4 \leqslant b_q \leqslant 11$.  Consequently, function \eqref{eq:cv_g} which is a nonnegative weighted sum of convex functions is also convex. The same statement (without any approximation) can be applied to show the convexity of the constraints under variable change of \eqref{eq:vc} (for some constraints, we need to apply an extra $\log$ to both sides of the inequality). To solve this problem, a relaxed continuous version of the proposed mixed-integer convex formulation is iteratively optimized in a loop \cite{boyd2007tutorial}. At each epoch, the continuous convex optimization is solved and obtained values for relaxed integer variables are rounded by a given precision. Then, we fix the acceptable rounded variables and solve the relaxed continuous convex problem again. The loop continues untill all the integer variables have valid values. The number of iterations is at most equal to (in practice, is usually less than) the number of integer variables. Furthermore, a simpler problem should be solved as the number of iteration increases because some of the integer variables are fixed during each loop.

\section{Numerical Results}\label{Sec_VI}
In this section, we use simulation results to demonstrate the performance of the convex formulation for TCS. The European Cost239 optical network is considered with the topology and traffic matrix given in \cite{khodakarami2014flexible}. Simulation constant parameters are $\abs{\beta_2}=20393$ $\text{fs}^2/\text{m}$ , $\alpha=0.22$ dB/km, $\mathcal{L}_{spn}=80$ km, $\nu=193.55$ THz, $n_{sp}=1.58$, $\gamma=1.3$ $1/\text{W/km}$, $\mathcal{F}=80$ MHz, $\varrho=113$ ps, $\sigma=14$ fs, $\mathcal{G}=20$ GHz, $\mathcal{B}=2$ THz, $\mathcal{P}_{trb}=36$ W, $\mathcal{P}_{edc}=3.2$ W, $\mathcal{P}_{fft}=4$ mW, $\mathcal{P}_{dsp}=3$ mW \cite{khodakarami2014flexible, arik2014adaptive}. We use MATLAB, YALMIP and CVX software packages for programming, modeling and optimization.
\begin{figure}
 \begin{minipage}{.48\textwidth}
\center{\includegraphics[scale=0.45]{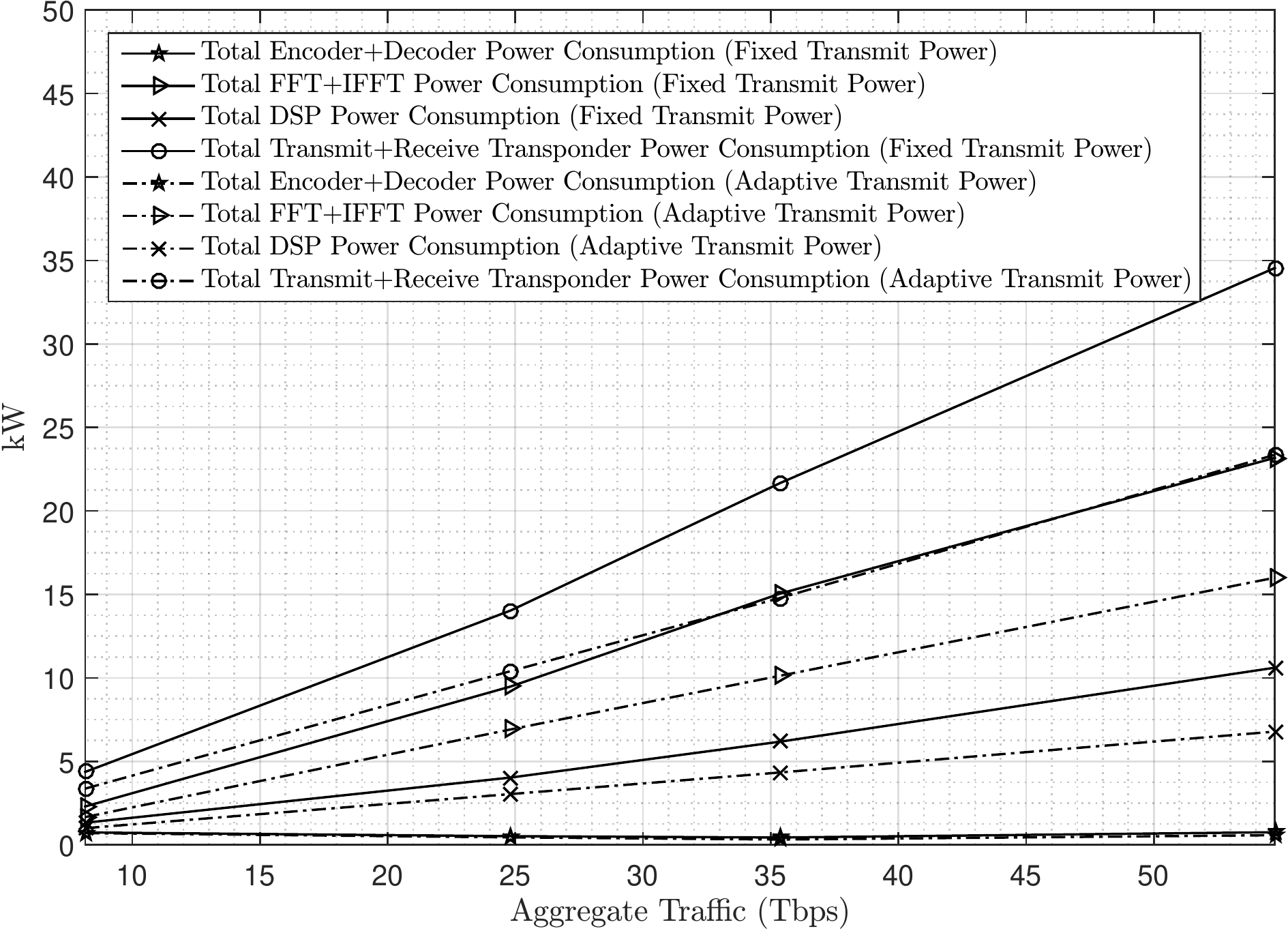}}
\center{\caption{\label{fig:powerAdapt} Total power consumption of different network elements in terms of aggregate traffic with and without adaptive transmit optical power assignment.}}
\end{minipage}
 \begin{minipage}{.48\textwidth}
\center{\includegraphics[scale=0.45]{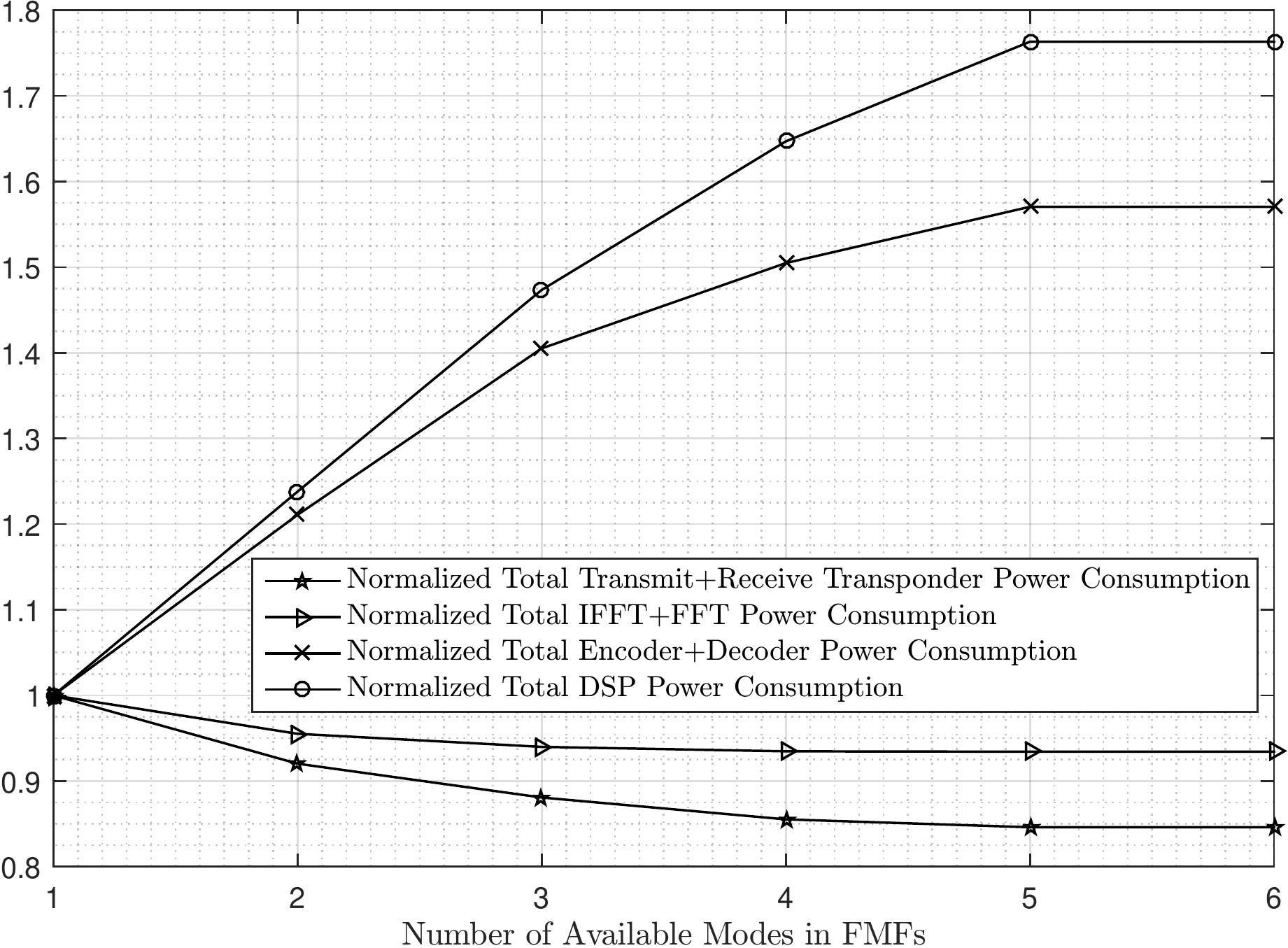}}
\center{\caption{\label{fig:modeAdapt} Normalized total power consumption of different network elements in terms of the number of available modes in FMFs.}}
\end{minipage}
 \begin{minipage}{.48\textwidth}
\center{\includegraphics[scale=0.45]{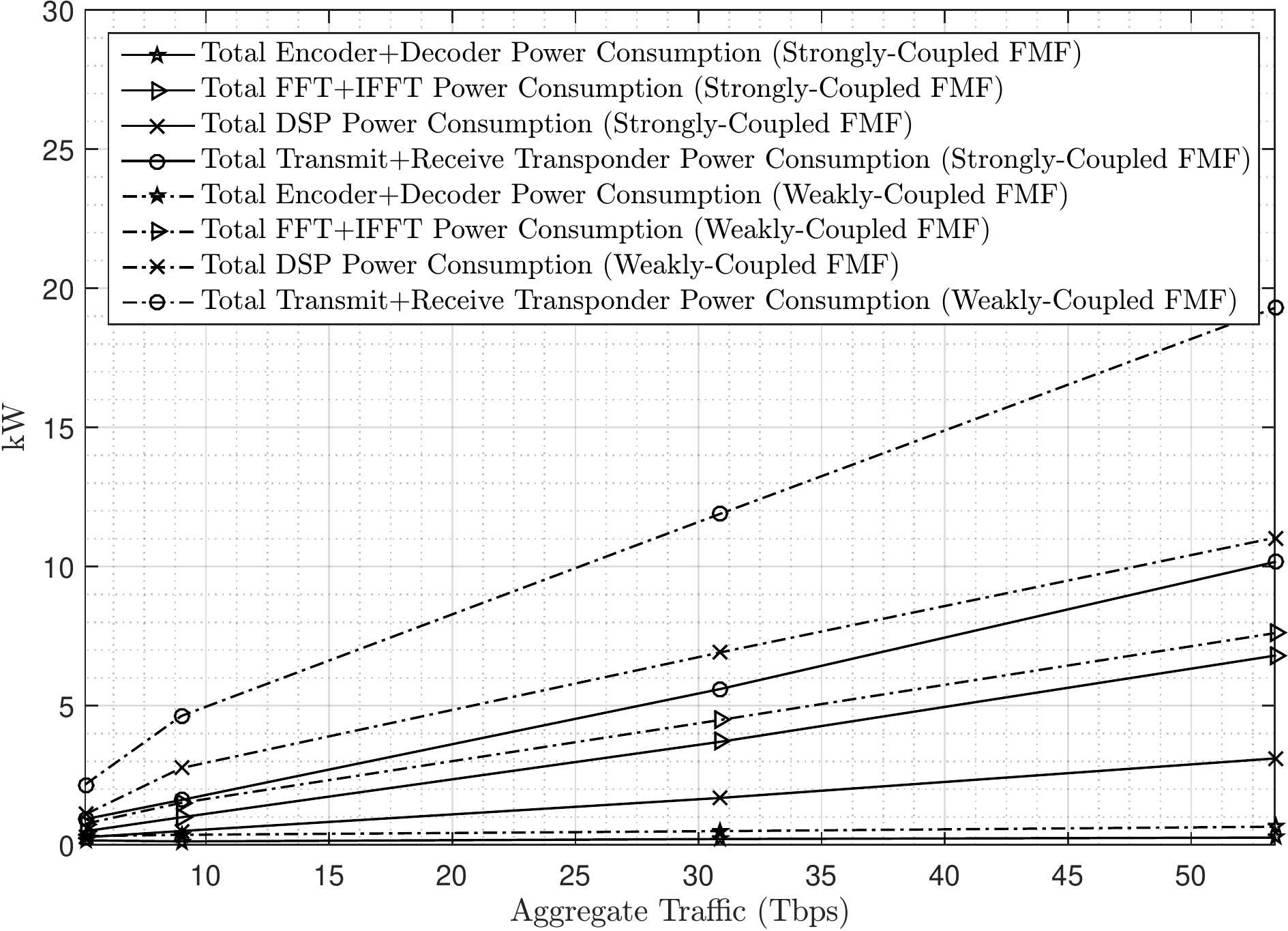}}
\center{\caption{\label{fig:fmfAssignment} Total power consumption of different network elements in terms of aggregate traffic for strongly- and weakly-coupled FMFs.}}
\end{minipage}
\end{figure}
The total power consumption of different network elements in terms of aggregate traffic with and without adaptive transmit optical power assignment has been reported in Fig. \ref{fig:powerAdapt}. We have used the proposed approach of \cite{gao2012analytical} for fixed assignment of transmit optical power. Clearly, for all the elements, the total power consumption is approximately a linear function of aggregate traffic but the slope of the lines are lower when transmit optical powers are adaptively assigned. As an example, adaptive transmit optical power assignment improves total transponder power consumption by a factor of $32\%$ for aggregate traffic of $60$ Tbps. Fig. \ref{fig:modeAdapt} shows total power consumption of different network elements versus number of available modes $\mathcal{M}$ in FMFs. The power consumption values are normalized to their corresponding values for the scenario with single mode fibers \ie $\mathcal{M}=1$. As $\mathcal{M}$ increases, the amount of transponder power consumption decreases but there is no considerable gain for $\mathcal{M} > 5$. Moreover, there is a tradeoff between DSP and FFT power consumption such that the overall transponder power consumption is a decreasing function of the number of available modes. Fig. \ref{fig:fmfAssignment} shows power consumption of different network elements in terms of aggregate traffic for strongly- and weakly-coupled FMFs. Obviously, total transponder power consumption is considerably reduced for strongly-coupled FMFs (in which group delay spread is proportional to square root of path lengths) in comparison to weakly-coupled FMFs (in which group delay spread is proportional to path lengths). This is the same as the results published in \cite{ho2013mode}. As an example, improvement can be more than $50\%$ for aggregate traffic of $60$ Tbps. Numerical outcomes also show that our convex formulation can be more than $20$ times faster than its mixed-integer nonlinear counterpart which is compatible with the results reported in \cite{hadi2017energy}.
\section{Conclusion}\label{Sec_VII}
Energy-efficient resource allocation and quality of service provisioning is the fundamental problem of green 3D FMF-based elastic optical networks. In this paper, we decompose the resource allocation problem into two sub-problems for routing and traffic ordering, and transponder configuration. We mainly focus on transponder configuration sub-problem and provide a convex formulation in which joint optimization of temporal, spectral and spatial resources along with optical transmit power are considered. Simulation results show that our formulation is considerably faster than its mixed-integer nonlinear counterpart and its ability to optimize transmit optical power can improve total transponder power consumption up to $32\%$. We demonstrate that there is a tradeoff between DSP and FFT power consumptions as the number of modes in FMFs increases but the overall transponder power consumption is a descending function of the number of available modes. We also calculate the power consumption of different network elements and show that  strongly-coupled FMFs reduce the power consumption of these elements.    
\bibliographystyle{IEEEtran}
\bibliography{Reference}

\end{document}